\documentclass[amssymb,aps,twocolumn,floats,showpacs]{revtex4-1}
\usepackage{color,graphicx,pstricks,float}
\usepackage{natbib}

\usepackage{fancybox}
\usepackage{epsfig}
\usepackage{graphicx}
\usepackage{tabularx}
\usepackage{ulem}
\usepackage{array} 

\usepackage{makecell}

\begin{document}

\title{Cluster mean field study of the Heisenberg model for CuInVO$_5$}

\author{Ayushi Singhania}
\author{Sanjeev Kumar}

\address{
Indian Institute of Science Education and Research (IISER) Mohali, Sector 81, S.A.S. Nagar, Manauli PO 140306, India \\
}

\begin{abstract}

Motivated by the experimental report of unusual low temperature magnetism in quasi one-dimensional magnet CuInVO$_5$, we present results of a cluster mean-field study on a spin-$1/2$ Heisenberg model with alternating ferromagnetic and antiferromagnetic nearest-neighbor coupling. We map out the ground state phase diagrams with varying model parameters, including the effect of an external magnetic field. An unexpected competition between different spin-spin correlations is uncovered. 
Multiple spin-flop transitions are identified with the help of component resolved correlation functions.
For the material-specific choice of model parameters we discuss the temperature dependence of specific heat and magnetic susceptibility, and compare our results with the available experimental data. 
A detailed account of spin-spin correlations allows us to present a microscopic understanding of the low-temperature magnetic ordering in CuInVO$_5$. 
Most notably, we identify the origin of an extra peak in the low temperature specific heat data of CuInVO$_5$ reported by Hase {\it et al.} \cite{Hase2016}.
\end{abstract}
\date{\today}


\maketitle

\noindent
\section{Introduction}

Spin-1/2 quasi-one-dimensional (Q1D) magnets are ideal candidates for observing fundamental quantum phenomena as the combination of low dimensionality and small spin-magnitude maximizes quantum fluctuations \cite{Vasiliev2018, Giamarchi2010}. This has motivated experimentalists for many decades to realize one dimensional quantum magnets \cite{Canepa2013, Arango2011}. These efforts have led to the discovery of many Q1D magnets and to the experimental verifications of various quantum phenomena \cite{Kinross}.
Indeed, quantum phase transitions driven by magnetic field or external pressure have been reported in low dimensional magnets such as TlCuCl$_3$, KCuCl$_3$, LiCuVO$_4$, CoNb$_2$O$_6$, etc. \cite{Oosawa2002, Tanaka2001,Nikuni2000, Oosawa2004, Christian2014,Banks2007,Cabrera2014,Ma2013}. Certain low-dimensional magnets have also been identified as being close to a quantum critical point \cite{Gavilano2005,Lake2005}.
Presence of extended quantum critical region has been inferred from the magnetic field dependence of excitations in copper pyrazine dinitrate \cite{Stone2003}.
Due to enhanced quantum fluctuations, Q1D magnets are also considered strong candidates for hosting quantum spin liquid states \cite{Nersesyan1996,Christian2014,Yoshida2015,Lecheminant2017}. Another aspect that makes low-dimensional magnets very interesting is the possibility of qualitatively new type of excitations \cite{Kohno2007,Skoulatos2017,Wang2018}. A classic example is that of spinon excitations in one-dimensional antiferromagnets \cite{Bera2017,Harrison1991}.  More recently a realization of longitudinal spin excitations, the so called Higgs mode, in certain Q1D magnets has been proposed \cite{Karbach2000,Kuroe2008,Coldea2010,Morris2014,Matsuda2015,Hase2016,Hase2017}. 

Recent experimental studies on spin-$1/2$ tetramer compound CuInVO$_5$ show unusual magnetism at low temperatures \cite{Hase2016}. Thermodynamic measurements, such as specific heat and magnetic susceptibility, show that a long-range ordered antiferromagnetic state exists below $2.7$K. There are two inequivalent Cu sites and the size of the ordered moment strongly differs at these two sites. This leads to a magnetization plateau in the magnetic field dependence at nearly half the saturation magnetization. While some of the features observed in CuInVO$_5$ can be explained within a simple mean-field approach, the presence of two peaks in the low-temperature specific heat and the presence of a cusp in the magnetic susceptibility remain as two of the unexplained features in the data \cite{Hase2016}. Furthermore, a microscopic picture of the ordered state and its evolution with magnetic field and temperature has been lacking.

Motivated by these puzzles in the experimental data on CuInVO$_5$, we present a comprehensive analysis of a four-sublattice one-dimensional Heisenberg model with three different nn exchange couplings. We make use of cluster mean-field (CMF) approach where intra-cluster interactions are treated exactly while inter-cluster interactions are treated at the mean-field level. The approach is well justified in the context of CuInVO$_5$ due to the existence of a hierarchy of coupling strengths as inferred from the experimental results \cite{Hase2016}. 
We find that treating inter-tetramer coupling beyond mean-field, which requires a minimum of $8$ sites in the cluster for the CMF study, brings out a subtle competition between two different spin-spin correlations. This emphasizes the presence of two distinct limiting phases in the model, and the ground state in CuInVO$_5$ is best understood as a compromise of these two  competing tendencies. Interestingly, the temperature dependence of the correlations is non-monotonic with certain spin-spin correlations strengthening with increasing temperature. 
Such effects are typically encountered in frustrated magnets where entropic effects at higher temperatures can help in enhancement of order \cite{Green2018,Champion2003, Guruciaga2016}. We also identify multiple 
spin-flop transitions in the presence of external field which highlight the inequivalence of spins within a tetramer. Most importantly, the subtle interplay between different spin-spin correlations accounts for the presence of an extra peak in the magnetic specific heat and a cusp in the magnetic susceptibility at low temperatures, in excellent agreement with the experimental data on CuInVO$_5$ \cite{Hase2016}.

The remainder of the paper is organized as follows. In Section II we define the model and discuss the CMF approach used for the study. Results are discussed in Section III where we begin by discussing the phase diagrams for the general choice of model parameters. This is followed by a discussion of various observables calculated for the parameters specific to CuInVO$_5$. For a clear understanding of the microscopic details we analyse the longitudinal and transverse spin-spin correlations between different pairs of spins. Summary and conclusions are presented in Section IV.

\noindent
\section{Model and Method} 
We begin with a Heisenberg model on a 1D chain of spin-$1/2$ tetramers in the presence of an external magnetic field. The model is described by the Hamiltonian,

\begin{eqnarray}
H & = &  \sum_{i = 1}^{N_t}[ J_2 ({\bf S}_{4i-3} \cdot {\bf S}_{4i-2} + {\bf S}_{4i-1} \cdot {\bf S}_{4i}) + J_1 {\bf S}_{4i-2} \cdot {\bf S}_{4i-1} \nonumber \\
  &   & + J_3 {\bf S}_{4i} \cdot {\bf S}_{4i+1}] - h_z \sum_{i=1}^{N_t} \sum_{j=0}^3 S^z_{4i-j}.
\label{Ham}
\end{eqnarray}

\noindent
Here, ${\bf S}_{4i-j}$ with $j = {0,1,2,3}$ are the Heisenberg spin operators belonging to the $i^{th}$ tetramer. 
$J_1 > 0 $, $J_2 < 0$, $J_3 > 0$ are the Heisenberg exchange constants
and $h_z$ is the magnitude of the applied magnetic field.
$N_t$ is the total number of tetramers, and periodic boundary condition is imposed via the identification ${\bf S}_{4N_t+1} \equiv {\bf S}_1$. For the analysis of the model Hamiltonian we will use $J_1 = 1$ as the elementary energy scale. This leaves us with $J_2$, $J_3$ and $h_z$ as free model parameters.
The inter-tetramer exchange $J_3$ is inferred to be much smaller than the intra-tetramer couplings $J_1$ and $J_2$ in CuInVO$_5$. 

In order to understand the nature of long-range magnetic order in the model Hamiltonian Eq. (\ref{Ham}), we employ the CMF approach. CMF method is an extension of the single-site  Weiss mean-field approximation, and has been very successful in studying the competition between different ordered states even in low dimensions \cite{Ren2014,Maier2005,Gotfryd2017}. 
It is well know that the Mermin-Wagner theorem prohibits the presence of any long range order at non-zero temperatures for isotropic spin Hamiltonians in dimensions $d \leq 2$ \cite{Mermin1966}. However, most low-dimensional magnets exhibit long-range order at small but finite temperatures \cite{Arango2011,Starykh2015,Dagotto1996}. CuInVO$_5$ is no exception to this trend as a long-range order sets in at $2.7$K. This apparent violation of Mermin-Wagner theorem can be understood in terms of the presence of magnetic anisotropies and/or the role of weaker inter-chain or inter-layer coupling. The importance of quantum effects in low-dimensional ordered magnets is typically reflected in the suppression of the ordered moment \cite{Balz2014}. 
The existence of long-range magnetic order in CuInVO$_5$ further justifies the use of CMF approach for describing low-temperature magnetism. 
One can argue that the mean-field aspect of the method is taking into account the three dimensional character of the magnetic system. Hence, the feature that CMF calculations lead to an ordered state at low enough temperatures is consistent with the experimental results.

\begin{figure}[t!]
\includegraphics[width=.98 \columnwidth,angle=0,clip=true]{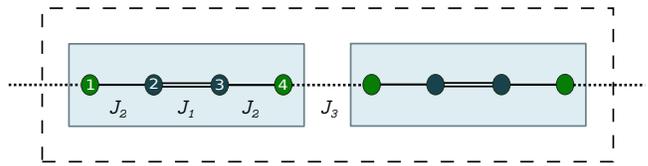}
\caption{(Color online) A schematic picture of the coupled tetramer model. Each dot represents a spin-$1/2$ and the nearest-neighbor couplings are indicated by double ($J_1$), solid ($J_2$) and dotted ($J_3$) lines.
}
\label{fig1}
\end{figure}

Although the CMF approach has been extensively discussed in literature \cite{Maier2005,Gotfryd2017}, for completeness, we briefly introduce the method here. Specifically, let us consider a one-dimensional system which can be thought of as repeated structure of clusters containing linear segments of $N_c$ spins. We want to treat the interactions within the cluster exactly while inter-cluster interactions will be treated approximately.
In a one dimensional system there are two edge spins, ${\bf S}_1$ and ${\bf S}_{N_c}$ that couple the central cluster to two adjacent clusters (see Fig \ref{fig1}). These two inter-cluster coupling terms can be approximated via the standard mean-field decoupling where ${\bf S}_i \cdot {\bf S}_{i+1}$ is replaced by $\langle {\bf S}_i \rangle  \cdot {\bf S}_{i+1} + {\bf S}_i \cdot \langle {\bf S}_{i+1} \rangle - \langle {\bf S}_i \rangle  \cdot \langle {\bf S}_{i+1} \rangle$ by ignoring the higher order fluctuation terms.
Therefore, the original Hamiltonian reduces to a cluster Hamiltonian in the presence of mean-fields that are experienced by the edge spins.
The mean fields acting on spins ${\bf S}_1$ and ${\bf S}_{N_c}$ are then calculated self-consistently.
For a cluster with $N_c$ spins of magnitude $1/2$, the size of the Hilbert space for the cluster Hamiltonian is $2^{N_c}$, and therefore the cluster Hamiltonian can be easily diagonalized exactly for $N_c \leq 12$. Note that in the general case where the mean fields are allowed to have components along $x$ and $y$ directions, the resulting mean-field Hamiltonian does not possess many of the symmetries of the full interacting Hamiltonian. Therefore, it is not generally possible to make use of symmetries to achieve diagonalizations of larger clusters. The quantum expectation values of the spin operators $\langle S^{\alpha}_i\rangle$ where $i$ denotes the site and $\alpha$ the spin  component, can be computed following the standard quantum statistical mechanics. The angular bracket denotes the quantum statistical average of the operator, and is defined for any operator $O$ as

\begin{eqnarray}
\langle O \rangle & = &  \frac{1}{{\cal Z}} Tr~[O~e^{-\beta H_c}],
\label{mf}
\end{eqnarray}

\noindent
where $\beta$ is the inverse temperature, $H_c$ is the cluster Hamiltonian, and ${\cal Z} = Tr ~ e^{-\beta H_c}$ is the partition function.
The process is repeated until a self-consistent solution is obtained upto a desired tolerance factor. In our calculations we take $10^{-5}$ as tolerance factor for convergence. 
As with all self-consistent approaches, we begin with a variety of initial mean-field configurations to ensure that the resulting self-consistent solution corresponds to a global minimum.

\noindent
\section{Results and Discussions} 

Before we consider the model parameters relevant to CuInVO$_5$, it is useful to explore the ground state phase diagram of the model in the parameter space $|J_2|/J_1$, $J_3/J_1$ and $h_z/J_1$. To obtain these CMF phase diagrams we work with an $8$-site cluster containing two tetramers. The justification for this choice will become clear in Sections III B. and III C. where we will present a comparison between results obtained using $4$-site and $8$-site clusters. 

\subsection{Spin-spin correlations in the ground state}

In order to characterize the ordered states at low temperature, we compute the transverse and longitudinal components of the spin-spin correlations defined by,

\begin{eqnarray}
C^{\perp}_{ij} & = &  \frac{1}{2} \langle S^+_{i} S^-_{j} + S^-_{i} S^+_{j} \rangle, \nonumber \\
C^{zz}_{ij} & = &  \langle S^z_{i} S^z_{j} \rangle.
\label{SCorr}
\end{eqnarray}
\noindent The total spin-spin correlations $C_{ij}$ can be obtained by adding the transverse and longitudinal components, $C_{ij} = C^{\perp}_{ij} + C^{zz}_{ij}$.

In the absence of external magnetic field, we present the evolution of total spin-spin correlations as a function of $|J_2|$ and $J_3$, keeping $J_1 = 1$ as the strongest exchange parameter. As expected, we find that $C_{23}$ retains its singlet-like character across the entire parameter regime covered in Fig. \ref{fig2} (see supplemental material). Similarly, $C_{12}$ and $C_{34}$ (see Fig. \ref{fig2}(c)) remain ferromagnetic in nature, except in the vicinity of the $J_2 = 0$ line where these correlations become vanishingly small. The behavior of $C_{23}$($C_{12}$/$C_{34}$) is not at all surprising since these spins are directly coupled via antiferromagnetic (ferromagnetic) interactions. Most interesting variation is noticed in $C_{14}$ and $C_{45}$. $C_{45}$ begins with a perfect singlet nature ($C_{45} \approx -0.75$) along $J_2 = 0$ line and the correlations diminish gradually as we move towards $J_3 = 0$ line. The behavior of $C_{14}$ is complementary to that of $C_{45}$. This can be easily understood as ${\bf S}_4$ can participate in only one perfect singlet, either with ${\bf S}_1$ or with ${\bf S}_5$. 
The tendency for singlet formation between ${\bf S}_4$ and ${\bf S}_5$ is easy to understand as these two spins are directly coupled via $J_3$. On the other hand, the singlet between ${\bf S}_1$ and ${\bf S}_4$ is mediated via an antiferromagnetic exchange $J_1$ and a ferromagnetic exchange $J_2$. 
The perfect singlet character for either pairs is disturbed when all the interaction strengths are finite. Instead, a compromise state with AFM correlations between both
${\bf S}_1$-${\bf S}_4$ and ${\bf S}_4$-${\bf S}_5$ pairs is preferred. It is important to note that this subtle competition is not captured in calculations that limit the cluster size to $4$-sites (single tetramer),  as in that case $C_{45}$ cannot be distinguished from $C_{14}$.
The correlation $C_{18}$ originates from the inter-cluster couplings where ${\bf S}_1$ and ${\bf S}_8$ belonging to the central cluster are coupled to mean-fields of ${\bf S}_8$ and ${\bf S}_1$, respectively. As expected, we find that this mean-field treatment restricts the correlation strengths to classical value of $-0.25$ (see Fig. \ref{fig2}(b)). 

\begin{figure}[t!]
\includegraphics[width=.98 \columnwidth,angle=0,clip=true]{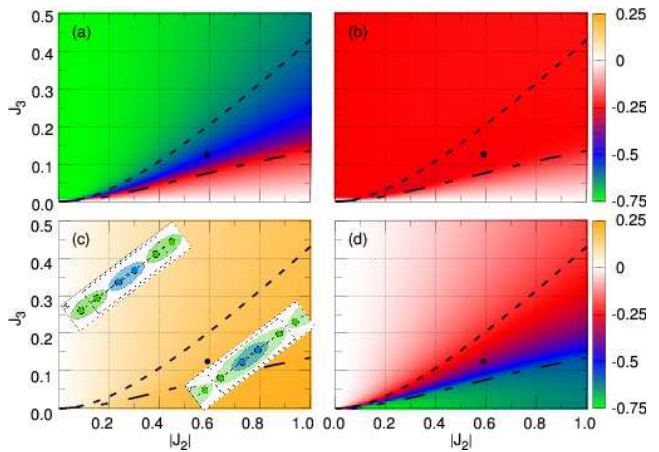}
\caption{(Color online) Variation of different spin-spin correlations with $|J_2|/J_1$ and $J_3/J_1$ for $h_z = 0$: (a) $C_{45}$, (b) $C_{18}$, (c) $C_{12}$ and (d) $C_{14}$. $C_{ij}$ are computed within CMF approach using $8$-site cluster. The dot indicates the location of the magnetic model for CuInVO$_5$ in the $|J_2|$-$J_3$ plane. The dashed line is an estimate for the path in parameter space where two different tendencies for singlet formation strongly compete (see text). In panel (c) we show the schematic picture of two limiting states.  The long-dashed line marks the separation between Neel-type long-range ordered state and the state consisting of non-interacting tetramers.
}
\label{fig2}
\end{figure}

The behavior of correlations between different spin pairs in the cluster points to the following three distinct ground states: (i) The simplest limit corresponds to $J_3 \rightarrow 0$ and $J_2 \rightarrow 0$ where the system is a collection of ${\bf S}_2$-${\bf S}_3$ singlets and isolated spins ${\bf S}_1$ and ${\bf S}_4$. (ii) If $J_3$ dominates over $J_2$, then the system can be considered close to a valance bond solid limit where two different type of singlets, one due to $J_1$ coupling and other due to $J_3$ coupling, are formed (see schematic picture near top-left corner in Fig. \ref{fig2}(c)). Of course, the exact singlet correlations are spoiled by the presence of the ferromagnetic $J_2$ coupling and also by the CMF treatment. As a consequence, an $\uparrow \uparrow \downarrow \downarrow$ type antiferromagnetic ordering with reduced magnetic moments emerges. Finally, (iii) in the case of $|J_2|$ dominating over $J_3$, the $C_{14}$ correlation achieve values close to that of perfect singlet, {\it i.e.}, $-0.75$, while $C_{45}$ is almost uncorrelated (compare Figs. \ref{fig2}(a),(d), and see schematic picture near bottom-right corner in Fig. \ref{fig2}(c)). By plotting the change in the self-consistent mean fields $\langle S^z_{1}\rangle$ and $\langle S^z_{8}\rangle$ as a function of $|J_2|$ for fixed values of $J_3$ (see supplemental material), we identify this limit in terms of the inequality $|J_2| > 8 J_3$, marked as long-dashed line in Fig. \ref{fig2}. The ground state in the region $|J_2| > 8 J_3$ corresponds to that of an isolated $4$-site cluster.
The magnetic phase diagram as inferred from the $C_{ij}$, therefore, consists of three qualitatively distinct regimes discussed above which are connected to each other continuously.

It is instructive to quantify the competition between different limiting cases. Fig. \ref{fig2}(a), (d) suggest that the key competition is between the singlet correlations $C_{14}$ and $C_{45}$. Solving the isolated 8-site cluster with open boundary condition, we find that the ground state energy is given by,
\begin{eqnarray}
E_{1} = -\frac{1}{4} (J_1 + 2 J_2 + 2 \sqrt{J_1^2 - 2 J_1 J_2  + 4 J_2^2 }).
\label{E1}
\end{eqnarray}
\noindent
On the other hand, the state in the limit $J_2 = 0$ is a collection of alternating singlets having energy per tetramer,
\begin{eqnarray}
E_{2} = -\frac{3}{4} (J_1 + J_3). 
\label{E2}
\end{eqnarray}
\noindent
Therefore, the competition between these two tendencies is strongest when the two energy contributions are equal. This gives us a relation between $J_2$ and $J_3$ which is obtained by numerically solving equations (\ref{E1}) and (\ref{E2}). The result is plotted as a dashed line in Fig. \ref{fig2}(a) and \ref{fig2}(d). The dot in Fig. \ref{fig2} represents the location of the magnetic model for CuInVO$_5$ in the parameter space of the model Eq. (\ref{Ham}). We note that CuInVO$_5$ is not far from this strongly competing regime, therefore, treating the $C_{45}$ correlations exactly is very important to capture the important aspects of magnetism in CuInVO$_5$.

\begin{figure}[t!]
\includegraphics[width=.98 \columnwidth,angle=0,clip=true]{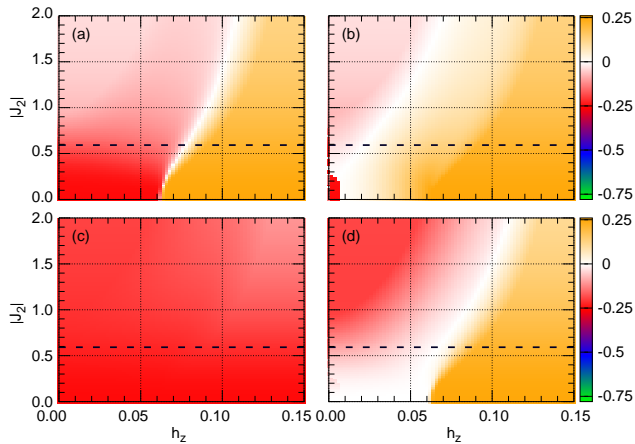}
\caption{(Color online) Variation of longitudinal component of spin-spin correlations with $|J_2|/J_1$ and $h_z/J_1$ for $J_3/J_1 = 0.125$: (a) $C^{zz}_{45}$, (b) $C^{zz}_{18}$, (c) $C^{zz}_{23}$ and (d) $C^{zz}_{14}$. Dashed horizontal lines correspond to the $|J_2|/J_1$ ratio estimated for CuInVO$_5$. 
}
\label{fig3}
\end{figure}

Next, we took at the dependence of spin-spin correlations on external magnetic field. In this case we discuss both the longitudinal and the transverse components of the correlations. For this purpose we fix the value of the inter-tetramer exchange $J_3 = 0.125$ and explore the phases in $h_z - J_2$ plane. 
The specific choice of the $J_3$ value is relevant to CuInVO$_5$ where $J_1$ and $J_3$ are estimated to be $240$K and $30$K, respectively \cite{Hase2016}.
For small values of $J_2$, the longitudinal and transverse components of $C_{23}$ are close to $-0.25$ and $-0.50$, respectively. These singlet-like correlations for $C_{23}$ remain unaffected by the external magnetic field in the regime $|J_2| < 1$. Interesting conclusions can be drawn by comparing the field dependence of component resolved $C_{18}$ and $C_{45}$. For small $J_2$, $C_{18}$ starts off with AFM correlations in the $z$-component and no correlations in the transverse direction, {\it i.e.}, $C^{zz}_{18} = -0.25$ and $C^{\perp}_{18} = 0$ (See Figs. \ref{fig3}(c) and \ref{fig4}(c)).
A sharp change in these correlations is found near $h_z = 0.01$ where the longitudinal component becomes close to zero and transverse component rises to $-0.25$. This is a clear signature of the spin-flop state involving a flopping of ${\bf S}_1$ and ${\bf S}_8$. The longitudinal component then gradually increases to positive values at the cost of reduction in transverse correlations in accordance with the standard picture of a spin-flop state evolving towards a canted state. Following the change in components of $C_{45}$ (say, at $J_2 = -0.58$ which is relevant for CuInVO$_5$) upon varying magnetic field highlights a similar effect for ${\bf S}_4$-${\bf S}_5$ pair. The transverse correlations reduce sharply near $h_z = 0.08$, and the longitudinal correlations vanish and then rapidly rise to positive values. Thus a clear picture emerges for the presence of two spin-flop transitions in this spin-1/2 tetramer model -- the first one corresponding to a flopping of edge spins and the second one to that of the central pair of spins. For still larger values of $h_z$, another spin-flop corresponding to ${\bf S}_2$-${\bf S}_3$ pair is present. Note that the anti-correlation between $C_{14}$ and $C_{45}$ is also present for finite magnetic fields (see panels (a) and (d) in Fig. \ref{fig3} and Fig. \ref{fig4}).


\begin{figure}[t!]
\includegraphics[width=.98 \columnwidth,angle=0,clip=true]{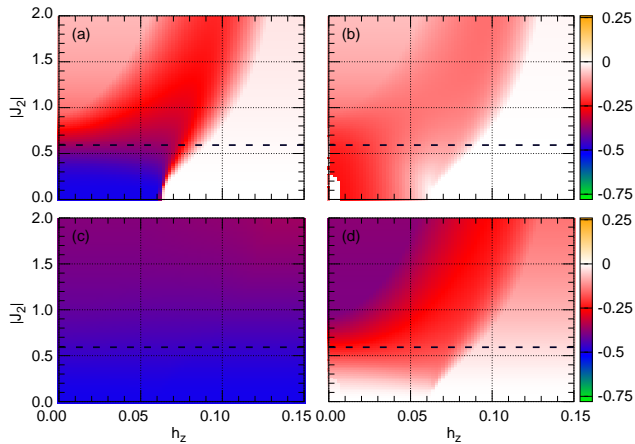}
\caption{(Color online) Variation of transverse component of spin-spin correlations with $|J_2|/J_1$ and $h_z/J_1$ for $J_3/J_1 = 0.125$: (a) $C^{\perp}_{45}$, (b) $C^{\perp}_{18}$, (c) $C^{\perp}_{23}$ and (d) $C^{\perp}_{14}$. Dashed horizontal lines correspond to the $|J_2|/J_1$ ratio estimated for CuInVO$_5$.
}
\label{fig4}
\end{figure}

Having discussed the broad picture for different spin-spin correlations and their component resolved evolution with magnetic field, we now focus on the parameter values considered relevant for CuInVO$_5$. We begin by discussing results for a 4-site cluster.

\subsection{Single-tetramer cluster}
\begin{figure}[t!]
\includegraphics[width=.98 \columnwidth,angle=0,clip=true]{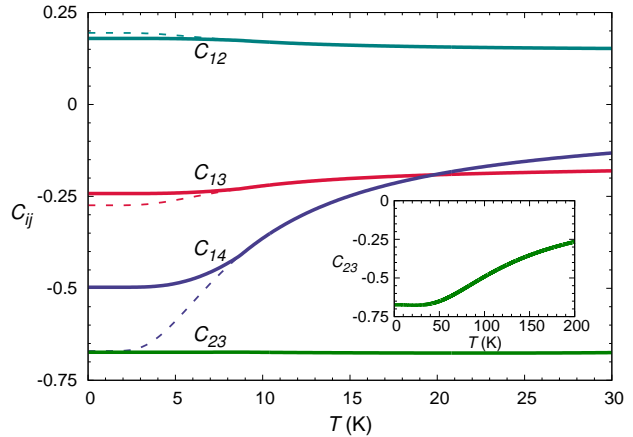}
\caption{(Color online) Spin-spin correlations $C_{ij}$ as a function of temperature for an isolated tetramer (dashed lines) and within CMF approach using a 4-site cluster (solid lines). Variation of $C_{23}$ over larger $T$ scale is shown in the inset.
}
\label{fig5}
\end{figure}

In this section, we will discuss results obtained via the CMF approach using a 4-site cluster. 
We begin by comparing the temperature dependence of spin-spin correlations obtained for an isolated tetramer and those via CMF with a $4$-site cluster. 
In the case of isolated tetramer, cluster is treated exactly with open boundary conditions where as in case of CMF, edge spins $\bf{S}_1$ and $\bf{S}_4$ couple to average fields $\langle \bf{S}_4 \rangle$, $\langle \bf{S}_1 \rangle$ respectively, via $J_3$.
Difference in the two sets of correlation functions vanish above $\sim 8$K. This indicates that the self-consistent mean-fields vanish above $8$K and the long-range order, which can be captured via CMF approach, is present below $8$K. Indeed, the main advantage of using a mean-field approach is to obtain results in thermodynamic limit. However, we point out a crucial shortcoming of the CMF approach applied to this system. The correlation $C_{14}$ for the two edge spins of a tetramer are treated better in an isolated tetramer. These correlation have a value, $C_{14} \approx -0.68$, close to that of a perfect singlet. In the mean-field approach the edge-spins are coupled to average fields due to finite $\langle {\bf S}_1 \rangle$ and $\langle {\bf S}_4 \rangle$, and therefore the correlations are strongly reduced. This can be observed for all the correlations involving the edge spins (see Fig. \ref{fig5}). The correlation of the central spin-pair $C_{23}$ is identical in the two calculations, as expected.

In addition to computing spin-spin correlation functions defined in Eq. (\ref{SCorr}), we also compute quantities that can be compared directly with the experiments.
To this end, we compute the specific heat and the magnetic susceptibility using the standard definitions,
\begin{eqnarray}
C_V(T) & = &  \frac{d\langle H \rangle}{dT}, \ \ \ \ \ \chi(T) = \frac{d\langle M_z \rangle}{dh}.
\label{CvChi}
\end{eqnarray}
We now present the comparison of specific heat calculations for isolated cluster and for the 4-site CMF approximation. For an isolated cluster the ground state belongs to the $S_T = 0$ sector and is characterized by singlet correlations between spin pairs ${\bf S}_1$-${\bf S}_4$ and ${\bf S}_2$-${\bf S}_3$.
This is indeed reflected in Fig. \ref{fig5} where the pair correlations $C_{23}$ and $C_{14}$ are found to be close to perfect singlet type. Treating the inter-tetramer interactions at the mean-field level spoils the singlet correlation $C_{14}$ as the edge spins now experience classical mean fields. The specific heat for an isolated cluster shows two broad peaks which can be naively associated with the loss of correlations $C_{14}$ at around $10$K, and the breaking of the stronger singlet between the central Cu spins at around $100$K. The CMF results lead to a sharp peak in $C_V$, signifying the on-set of long-range order below $\sim 8$K.

\begin{figure}[t!]
\includegraphics[width=.98 \columnwidth,angle=0,clip=true]{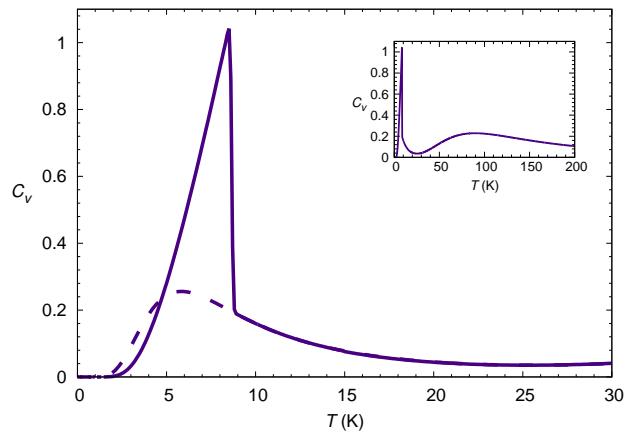}
\caption{(Color online) Specific heat as a function of temperature for isolated tetramer (dashed line) and for CMF approximation with 4-site cluster (solid line). 
The inset shows the behavior over wider temperature scale for the CMF approximation.
}
\label{fig6}
\end{figure}

\begin{figure}[t!]
\includegraphics[width=.98 \columnwidth,angle=0,clip=true]{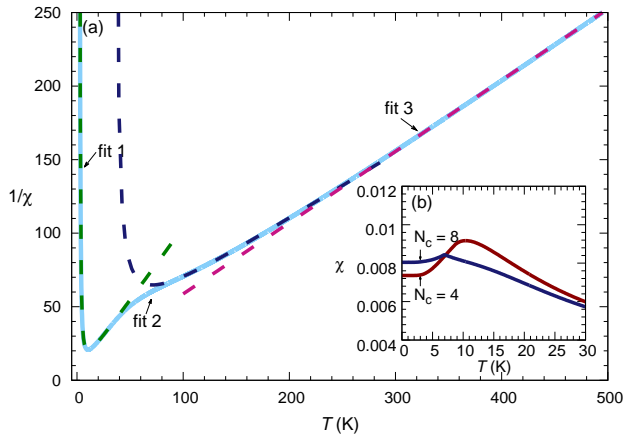}
\caption{(Color online) Inverse magnetic susceptibility, $\chi^{-1}$, as a function of temperature for an isolated tetramer. The dashed lines are the best fits corresponding to three different temperature regimes (see text). Inset shows the result for $\chi(T)$ within CMF approximation using $4$-site and $8$-site clusters.
}
\label{fig7}
\end{figure}

In order to confirm the simple picture proposed from the spin-spin correlation and the specific heat calculations, we now show the magnetic susceptibility results. 
If the simple picture of two-step loss of correlations is indeed true then it should have specific consequences for the behavior of magnetic susceptibility. 
To verify this, we plot the inverse magnetic susceptibility obtained for an isolated cluster in Fig. \ref{fig7}.
Given the tendency for singlet formation at low temperatures, we fit the magnetic susceptibility differently in three temperature regimes. In the range $0$K $< T < 40$K, we fit the susceptibility via the following behavior for singlets \cite{Buschow1997} (see supplemental material),
\begin{equation}
    \chi(T) = \frac{a_1}{T} \frac{e^{(-b_1/T)}}{1+3 e^{(-b_1/T)}}.
\end{equation}

\noindent
In the above, the fitting parameter $a_1$ contains information about number of singlets, and $b_1$ is related to the excitation gap. 
In the regime $40$K $< T < 300$K, the system should display a mixed behavior since the weaker singlets cease to exit and the participating spins will now contribute as free paramagnetic moments. Therefore, we fit the susceptibility via,
\begin{equation}
  \chi(T) = \frac{a_2}{T} \frac{e^{(-b_2/T)}}{1+3 e^{(-b_2/T)}}  + \frac{c_2}{T-d_2}.
\end{equation}

\noindent
The second term is simply Curie-Weiss behavior and the two fitting parameters contain information regarding the total number of paramagnetic moments and the Curie-Weiss temperature.
In the high-temperature regime, one expects a total Curie-Weiss behavior for all the constituent spins. Therefore, a Curie-Weiss fit, $ \chi(T) = \frac{c_3}{T-d_3}$, is used in the range $300$K $< T < 600$K. The actual $\chi^{-1}(T)$ and the three fits discussed above are shown in Fig. \ref{fig7}.
From the quality of the fit the following simple picture is reconfirmed. At low temperature, the magnetic susceptibility fits very well to a singlet behavior. At intermediate temperatures, two of the spins get free and contribute to Curie-Weiss susceptibility. Finally a paramagnetic behavior emerges at high temperatures. The obtained fit parameters differ slightly from the above picture in terms of number of spins contributing to susceptibility as singlets or paramagnetic moments at different temperatures (see supplemental material). 

We find that while the tendency for singlet formation below $100$K between ${\bf S}_2$ and ${\bf S}_3$ and the long-range order to a Neel state with $\uparrow \uparrow \downarrow \downarrow$
pattern below about $10$K is obtained within the $4$-site CMF approach, the experimental observation of a second peak in the specific heat at about $2.7$K is not consistent with the CMF results. We argue that treating inter-tetramer interactions beyond mean-field is the key to understanding the magnetism of CuInVO$_5$. We discuss the 8-site CMF results in the next section. Nevertheless, we already find that $8$-site CMF results for magnetic susceptibility are qualitatively different from those obtained for $4$-site CMF (see inset in Fig. \ref{fig7}). A cusp-like feature followed by a broad hump is reported in the experiments which seems to be captured within 8-site CMF calculations.
Clearly, if the interaction $J_{3}$ happens to be stronger than the ferromagnetic interaction $J_2$ then the system would prefer to form singlets between ${\bf S}_2$ and ${\bf S}_3$ and ${\bf S}_4$ and ${\bf S}_5$ instead of a pair of singlets within a tetramer. In fact, even if $J_3$ is much smaller than $J_2$, since $J_3$ is antiferromagnetic in nature it may be important to retain the correlations in the inter-tetramer interaction. The simplest way to achieve this is to increase the cluster size to $8$-spins (two-tetramers) where one central inter-tetramer exchange term will be treated exactly. Next, we present results on CMF using two- and three-tetramer units as the cluster.

\subsection{Beyond single-tetramer cluster}

\begin{figure}[t!]
\includegraphics[width=.98 \columnwidth,angle=0,clip=true]{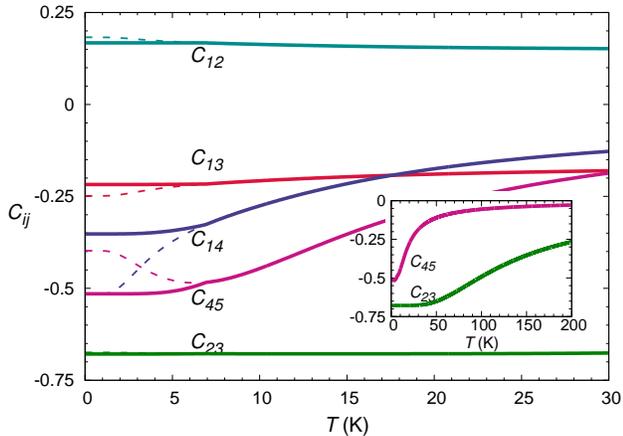}
\caption{(Color online) Spin-spin correlations, $C_{ij}$, as a function of temperature for an isolated 8-site cluster (dashed lines) and within the CMF approach using an 8-site cluster (solid lines). Variations in $C_{23}$ and $C_{45}$ over a wider $T$ range is shown in the inset.
}
\label{fig8}
\end{figure}

\begin{figure}[t!]
\includegraphics[width=.98 \columnwidth,angle=0,clip=true]{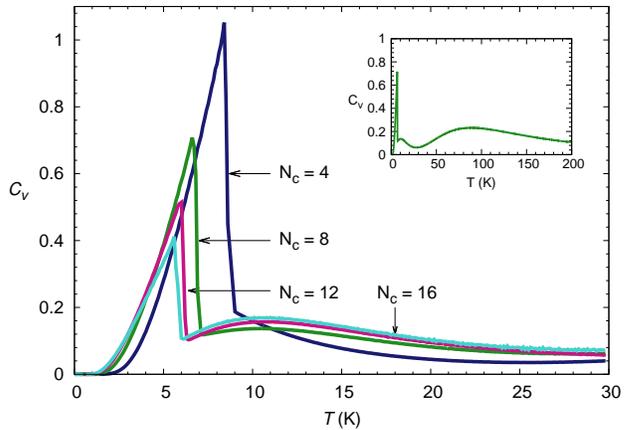}
\caption{(Color online) Specific heat as a function of temperature within the CMF approximation for $4$-site, $8$-site, $12$-site and $16$-site clusters. The behavior across a broader temperature scale is displayed in the inset.
}
\label{fig9}
\end{figure}

We begin by presenting the spin-spin correlation functions for different pairs as a function of temperature. Note that the most important correlation that was missing in the $4$-site cluster treatment is $C_{45}$. Exact solution of the isolated $8$-site cluster shows that at $T=0$,  $C_{14}$ is antiferromagnetic in nature and larger in magnitude than $C_{45}$. With increasing temperature $|C_{14}|$ reduces rapidly (see Fig. \ref{fig8}). Interestingly, this decrease of $|C_{14}|$ is accompanied by an increase of $|C_{45}|$.
Note that it is rather unusual to find an increase in the magnitude of correlations as a function of temperature. This hints towards competing tendencies for order in the ground state. We can comprehend this finding as follows.
Spin ${\bf S}_4$ can have singlet type correlations with ${\bf S}_5$ due to the antiferromagnetic exchange constant $J_3$. However, it can also have quantum antiferromagnetic correlations with spin ${\bf S}_1$ due to combined effect of an antiferromagnetic $J_1$ and ferromagnetic $J_2$. These two tendencies for singlet correlations are competing in the ground state, and for the material-specific values of the exchange parameters a dominant antiferromagnetic correlations with spin ${\bf S}_1$ is energetically favored. With increasing temperature, a weakening of longer-range correlations ($C_{14}$) allows for strengthening of $C_{45}$.
This intriguing interplay of two competing tendencies for singlet formation is apparent in our discussion of the model for generic parameter values (compare Fig. \ref{fig2}(a) and Fig. \ref{fig2}(d)). Interestingly, this competition between different singlet choices is also at play when temperature varies, and has
consequences for physical observables. The fact that different spin-spin correlations are being affected at different temperatures should be reflected in specific heat results. To verify this we plot in Fig. \ref{fig9} the specific heat calculated within the CMF approach using $4$, $8$, $12$ and $16$ site clusters. In contrast to the results for $4$ site cluster, two peaks at low temperatures are found in the $8$, $12$ and $16$ site CMF calculations. 


\begin{figure}[t!]
\includegraphics[width=.98 \columnwidth,angle=0,clip=true]{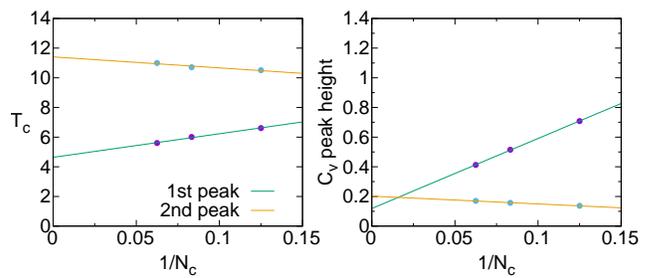}
\caption{(Color online) Scaling of peak locations, $T_p$, and peak values, $C_V(T_p)$, in the specific heat shown in Fig. \ref{fig9} with inverse cluster size.
}.
\label{fig10}
\end{figure}

The results suggest that the most important improvement to the $4$-site CMF results already occurs when we use $8$ site cluster and hence treat inter-tetramer interaction exactly.
The relative strength and position of the two low-temperature peaks in $C_V$ change as we increase the cluster size (see Fig. \ref{fig10}).
The first peak which is related to the long-range order reduces with increasing system size. Although, the scaling based on 3 data points is not conclusive, the estimates for the peak locations $T_p$
obtained from the extrapolated data are in very good agreement with the experiments with an overestimations of about $1.5$K.

More importantly, it is ruled out that any new peaks in the specific heat arise with adding more tetramers to the cluster used in the CMF approach. Note that the experimental plot for $C_V$ also contains contribution from phonons which needs to be subtracted in order to identify the pure magnetic contribution. While the phonon contribution will mask the high temperature peak around $90$K (see inset in Fig. \ref{fig9}), the two lower temperature peaks are easily identified in the experimental data \cite{Hase2016}. 

The magnetic field dependence of the spin-spin correlations is already discussed in Figs. \ref{fig3} and \ref{fig4} for generic choice of model parameters. In order to obtain the results specific to CuInVO$_5$ we simply need to find the appropriate values of model parameters. These results were obtained for $J_3/J_1 = 0.125$, a ratio motivated from the estimated values of $J_1$ and $J_3$ in CuInVO$_5$. In the material, $|J_2|/J_1$ is estimated to be $0.58$ and we can focus on the $|J_2| = 0.58$ line to discuss the field dependence of correlations in CuInVO$_5$. A partial spin-flop is present at low magnetic fields which leads to a magnetization plateau at $h_z = 0.08 J_1$ which turns out to be around $30$T when appropriate conversion factors are included. This coincides very well with the presence of the plateau in the field dependence of magnetization (see Fig. 5 in \cite{Hase2016}). If the simple picture of partial spin-flop transition is valid, then we should see further increase in magnetization at yet higher magnetic fields. Indeed, we obtain saturation magnetization at about $145$T (see supplemental material).

\begin{figure}[t!]
\includegraphics[width=.98 \columnwidth,angle=0,clip=true]{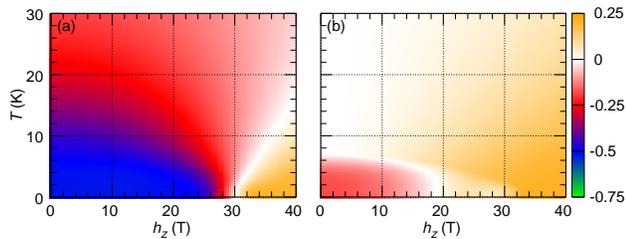}
\caption{(Color online) Temperature and field dependence of, (a) spin-spin correlations $C_{45}$ and (b) product of mean-field variables, $\langle {\bf S}_1 \rangle \cdot \langle {\bf S}_8 \rangle$ , for the parameter values specific to CuInVO$_5$. Temperature (magnetic field) axis is in the physical unit Kelvin (Tesla).
Note that the quantity plotted in (b) is finite for a long-range ordered magnetic state.
}
\label{fig11}
\end{figure}

Combining the results on temperature and magnetic field dependence of mean-field parameters and spin-spin correlations, we present a $h_z - T$ phase diagram in
Fig. \ref{fig11}. The dot product of mean fields $\langle {\bf S}_1 \rangle \cdot \langle {\bf S}_8 \rangle$ is a measure of the long-range order in the system. As we can clearly see in Fig. \ref{fig11}(b), for small values of field there is a transition close to $5$K from a long-range ordered to disordered state. However, even in the disordered state there are certain short-range correlations that remain finite. The most important of these is $C_{45}$ which is shown in Fig. \ref{fig10}(a). These correlations remain finite upto larger temperatures and show a significant variation near $T = 10$K. This variation is the underlying reason for a broad peak in the specific heat near $10$K. The evolution of mean-field variables with magnetic field shows that the edge spins gradually approach an aligned state starting with an anti-aligned state. The saturation alignment is achieved at about $30$T. Note that while the edge spins are aligned, the central spins still retain considerable singlet correlations and therefore the contribution to magnetization is from these edge spins leading to the magnetization plateau in the experimental data \cite{Hase2016}.

It is possible to further improve the mean-field description of the model by using different extensions of CMF approach. Two such extensions are correlated CMFT and quantum correlated CMFT \cite{Yamamoto2009,Zimmer2014,Zimmer2016}. However, the most important aspect of the magnetic model for CuInVO$_5$ is already captured by our minimal description where the inter-tetramer interaction is included in exact manner. While some of the quantitative details, such as the relative magnitude of the low-temperature peaks, the exact location in temperature of the peaks, etc., are likely to change in a more accurate treatment of the model, the qualitative character is well described in our CMF approach.

\noindent
\section{Summary and Conclusion} 
We have performed cluster mean field analysis of a one-dimensional Heisenberg model with alternating signs of exchange constants. The choice of the model is motivated by the unusual low-temperature magnetism in CuInVO$_5$ \cite{Hase2016}. We map out the nature of spin-spin correlations as a function of different model parameters. The results are obtained via CMF approach with an $8$-site cluster which, in contrast to the $4$-site cluster study \cite{Hase2016}, captures the effect of the inter-tetramer coupling beyond mean-field. It turns out to be an essential ingredient for understanding some of the experimental observations, in particular, multiple peaks in the low-temperature specific heat.
Due to a better treatment of quantum correlations of the inter-tetramer coupling, an interesting competition between two qualitatively different ground states is uncovered. These ground states are best understood in the limiting cases $J_3 \rightarrow 0$, and $J_2 \rightarrow 0$.
In the limit $J_3 \rightarrow 0$ the system is a collection of isolated tetramers and the ground state for an isolated tetramer is characterized in terms of 
quantum antiferromagnetic correlations between spins ${\bf S}_2$ and ${\bf S}_3$ and those between ${\bf S}_1$ and ${\bf S}_4$. The latter of these relies on the ferromagnetic exchange $J_2$ as a mediator. On the other hand, in the limit $J_2 \rightarrow 0$ the ground state becomes a collection of alternating singlets, one mediated by exchange $J_1$ and other by $J_3$. However, this state is only accessible when quantum correlations of the inter-tetramer interactions are retained. 
When $J_2$ and $J_3$ are both finite, a competition between these qualitatively distinct states is realized. Our study shows that the ground state of the CuInVO$_5$ emerges out of this competition. The above description of the low-temperature magnetism in CuInVO$_5$ is inferred from our analysis of the model for material-specific values of the parameters. We show that an interesting evolution of the competition between different spin-spin correlations exists not only with variation of model parameters but also with increasing temperature. Correlations for certain pair of spins even increase with increasing temperature which is contrary to the general expectations that thermal effects reduce the correlations. Magnetic susceptibility calculations further allow us to identify three distinct regimes in temperature corresponding to a complete paramagnetic behavior at high temperature, a singlet-like behavior at low-temperatures, and a mixed behavior at intermediate temperatures. At intermediate temperatures some of the spins get free from singlets while other retain strong singlet correlations.
This is consistent with the experimental finding of the magnetization plateau at nearly half the saturation magnetization. By tracking transverse and longitudinal spin-spin correlations, we observe a two-step spin-flop transition in the model. 
The most important implication of this competition of correlations captured in our CMF study is the existence of multiple peaks in the specific heat -- a puzzling feature reported in the experimental data on CuInVO$_5$ \cite{Hase2016}. 


\noindent
\section{Acknowledgments}
We acknowledge the use of High-Performance Computing Facility at IISER Mohali.

%

\widetext
\clearpage

\setcounter{equation}{0}
\setcounter{figure}{0}
\setcounter{table}{0}
\setcounter{page}{1}
\makeatletter

\begin{center}
	\textbf{\large Supplementary Material}
\end{center}

	\renewcommand{\thefigure}{S\arabic{figure}}
	\maketitle
	
	\section*{Spin-spin correlations}
	\begin{figure}[ht]
		\centering
		\includegraphics[scale=0.85]{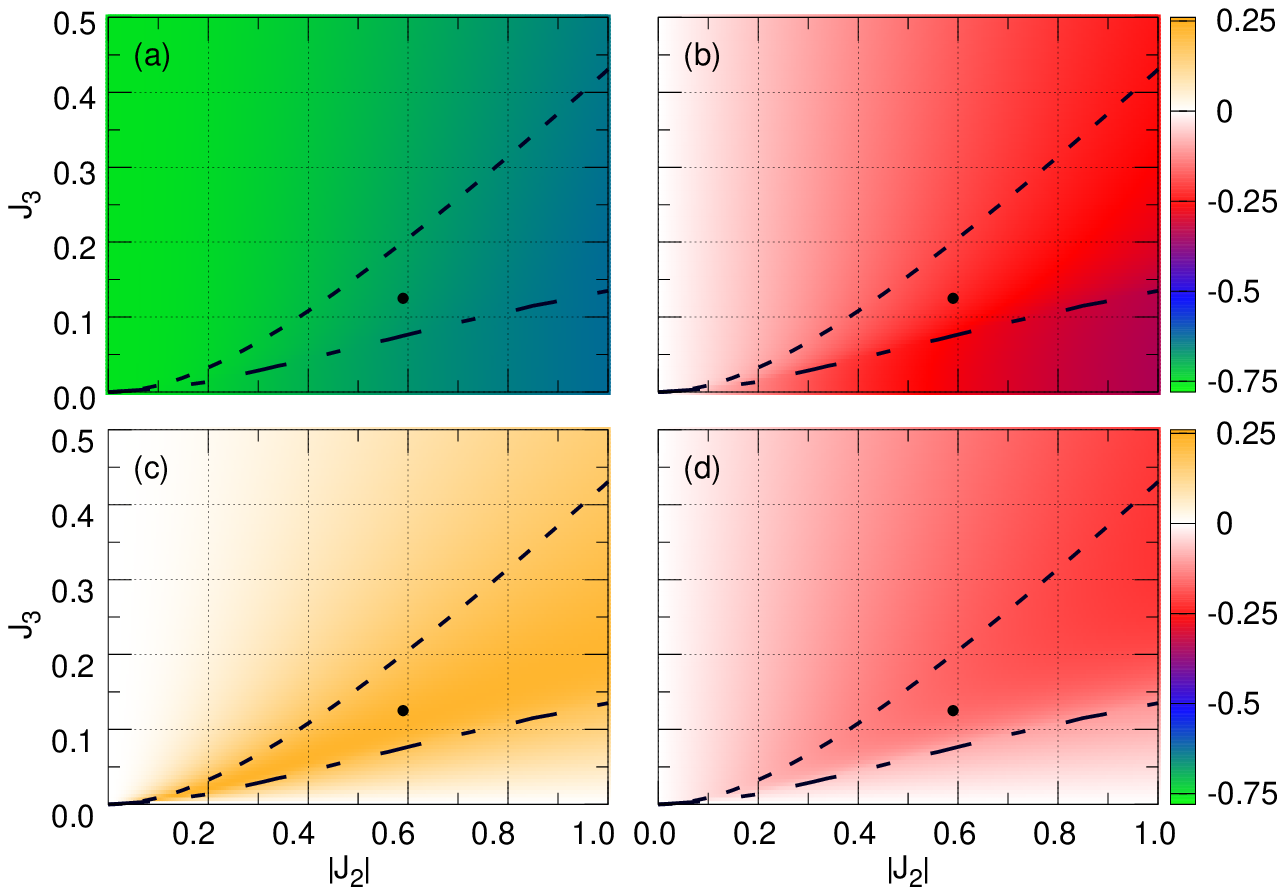}
		\caption{Variation of different spin pair correlations $(C_{ij})$ with $|J_2|/J_1$ and $J_3/J_1$ for $h_z =0$ computed within CMF for 8-site cluster: (a) $C_{23}$, (b) $C_{13}$, (c) $C_{15}$ and (d) $C_{35}$. The dashed and long-dashed lines are identical to those shown in Fig. 2 in main text. The dot represents the location on CuInVO$_5$ in the parameter space.}
		\label {fig1}
	\end{figure}
	In continuation of the discussion in Section III A about three different limiting states in $J_2-J_3$ parameter space, we present other relevant spin pair correlations in Fig. S1. $J_1$ being the strongest exchange parameter, \textbf{$S_2$-$S_3$} pair retains its strong-singlet character throughout the parameter space (see Fig. \ref{fig1} (a)). However, as expected, the correlation begins to weaken as $|J_2|/J_1$ approaches $1$. Correlation between \textbf{$S_1$} and \textbf{$S_3$} increases with increasing $|J_2|$ (see Fig. \ref{fig1} (b)), which is also related to the weakening of $S_2-S_3$ singlet. This is where $J_2$ starts competing with $J_1$. Ordering of spins is largely controlled by $C_{45}$ correlation (see Fig. 2(a) in main text), its affect can also be seen in $C_{15}$ and $C_{35}$ (see Fig. S1 (c)-(d)). Correlation between $S_1$ and $S_5$ cease to exist in $J_3 \Rightarrow 0$ and $J_2 \Rightarrow 0$, however it changes continuously in the intervening region. This correlation is mediated via $J_3$, as singlet strength begins to increase $C_{35}$ decreases. A strong crossover is observed around $J_3 \sim 0.2|J_2|$ for $C_{35}$, this is the region when $C_{45}$ varies from classical anti-parallel correlation to quantum mechanical singlet-like bond characterized by values less than $-0.25$.   
	
	\begin{figure}[ht]
		\centering
		\includegraphics[scale=0.7]{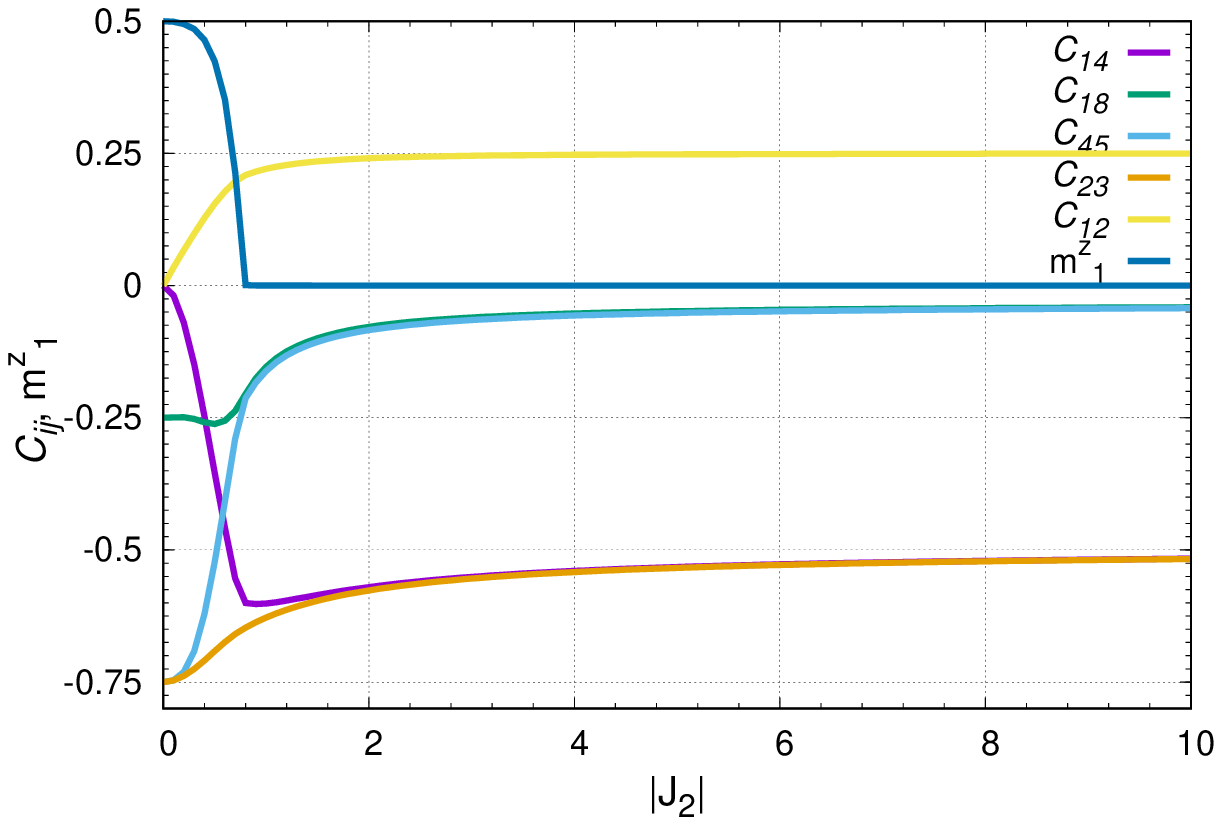}
		\caption{Variation of spin-spin correlations and mean fields $m_1^z$ as a function of $|J_2|$ for $J_3 = 0.1$.}
		\label{fig2}
	\end{figure}
	
	We also show the evolution of correlation functions and self-consistent mean field $m^z_1$ in the limit $|J_2| >> J_3$ in Fig. \ref{fig2}. The mean field vanishes beyond $|J_2| \sim 0.8$, where the 8-site cluster behaves like two weakly coupled tetramers (see $C_{45}$ in Fig. \ref{fig2}). For $|J_2| < 0.8$, $\bf{S}_4$ continuously forms strong antiferromagnetic correlation with $\bf{S}_1$ at the cost of the singlet bond with $\bf{S}_5$. Similar calculations for other values of $J_3$ show that one can define $|J_2| = 8 J_3$ as the line separating the Neel-type long-range ordered states from that consisting of weakly coupled tetramers schematically shown in Fig. 2(c) in main text.

	\section*{Response to magnetic field}
	\begin{figure}[ht]
		\centering
		\begin{minipage}[b]{0.45\textwidth}
			\includegraphics[width=\textwidth]{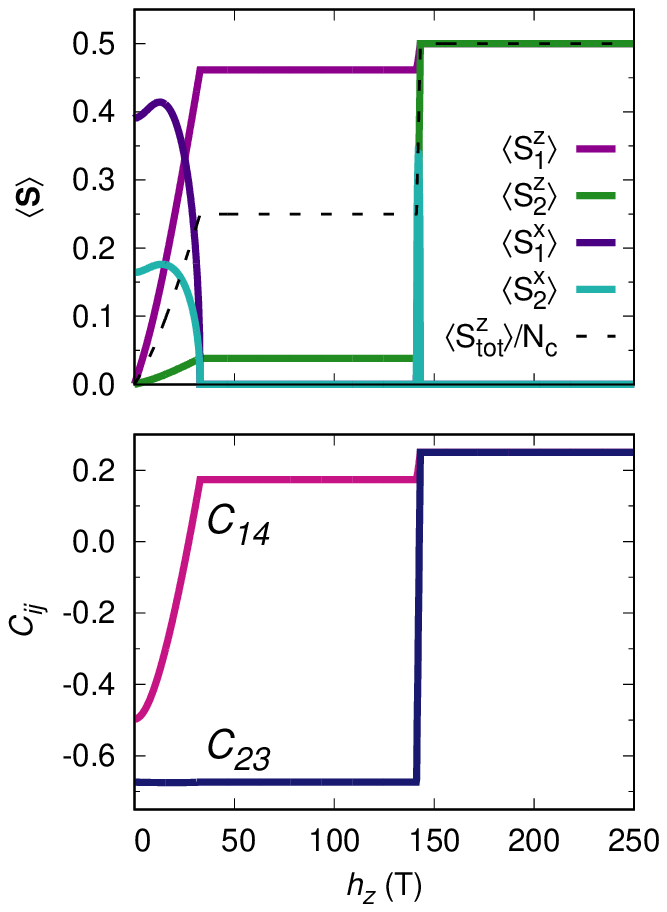}\\
			(a)
		\end{minipage}
		\hfill
		\begin{minipage}[b]{0.45\textwidth}
			\includegraphics[width=\textwidth]{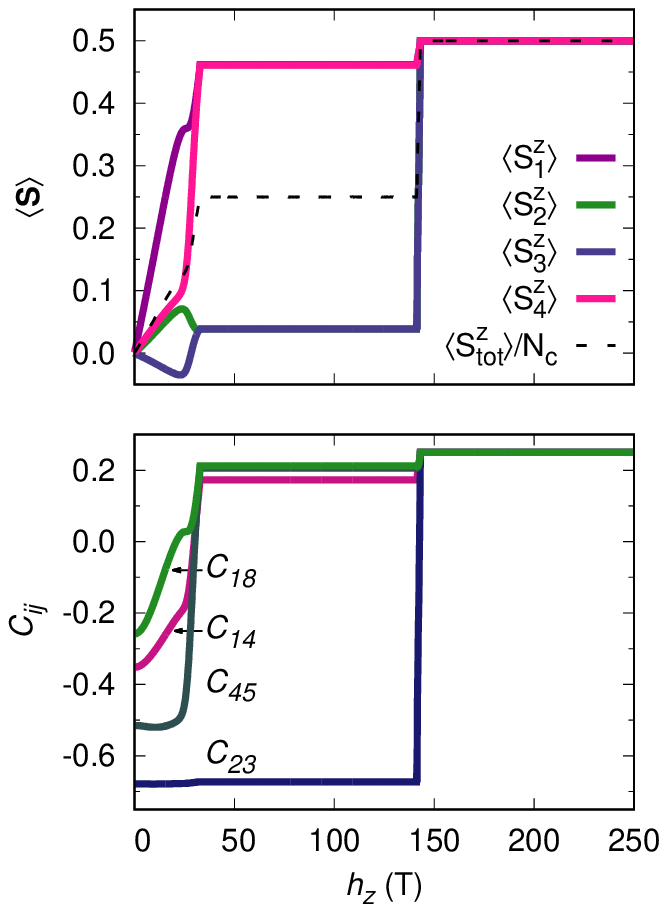}\\
			(b)
		\end{minipage}
		\caption{Average spin magnetic moment (upper panels) as a function of applied magnetic field and corresponding behaviour of correlations (lower panels) computed using CMF at $T=0.01$K for (a) single tetramer and (b) two-tetramer cluster.}
	\end{figure}
	Multiple spin flops are inferred from $h_z-J_2$ phase diagram of 8-site cluster, as discussed in the main text (Figs. 3, 4). For the material specific value of $J_2$, we observe a spin flop at around $h_z \sim 0.08$ and a final re-orientation of spins at higher fields. Figs. S3(a), S3(b) show a component resolved magnetic moment variation with applied field. For CMF results using a single tetramer cluster, we find a two step saturation of magnetic moments. A spin flop transition to a direction perpendicular to applied field is observed, which is followed by the first magnetization $(M = g \mu_B \langle S \rangle)$ plateau around $\sim 30 T$ (follow the black dashed lines Fig. S3). The first plateau is related to the loss of \textit{$C_{14}$} correlation whereas the full saturation of magnetic moments takes place when the singlet between $\mathbf{S_2}$ and $\mathbf{S_3}$ breaks, which is clearly visible in the correlation plot. Two-step saturation of magnetization is also consistent in 8-site (two tetramer) cluster calculations. However, the details of the field dependence are slightly different. In Fig. S3(b), average moments show a non-linear increase below the first magnetization plateau. This is different from single tetramer results where average magnetic moment in the direction of field shows a linear increase. The non-linear increase in directly related to the loss of \textit{$C_{14}$} and \textit{$C_{18}$} correlations. The experimentally observed behavior is indeed non-linear and is consistent with the 
	results obtained using two-tetramer CMF. This further highlights the importance of treating the inter-tetramer coupling beyond mean-field for an improved description of the experimental data.   \\

	\section*{Fitting details of susceptibility for single-cluster}
	\begin{figure}[ht]
		\centering
		\includegraphics[scale=0.65]{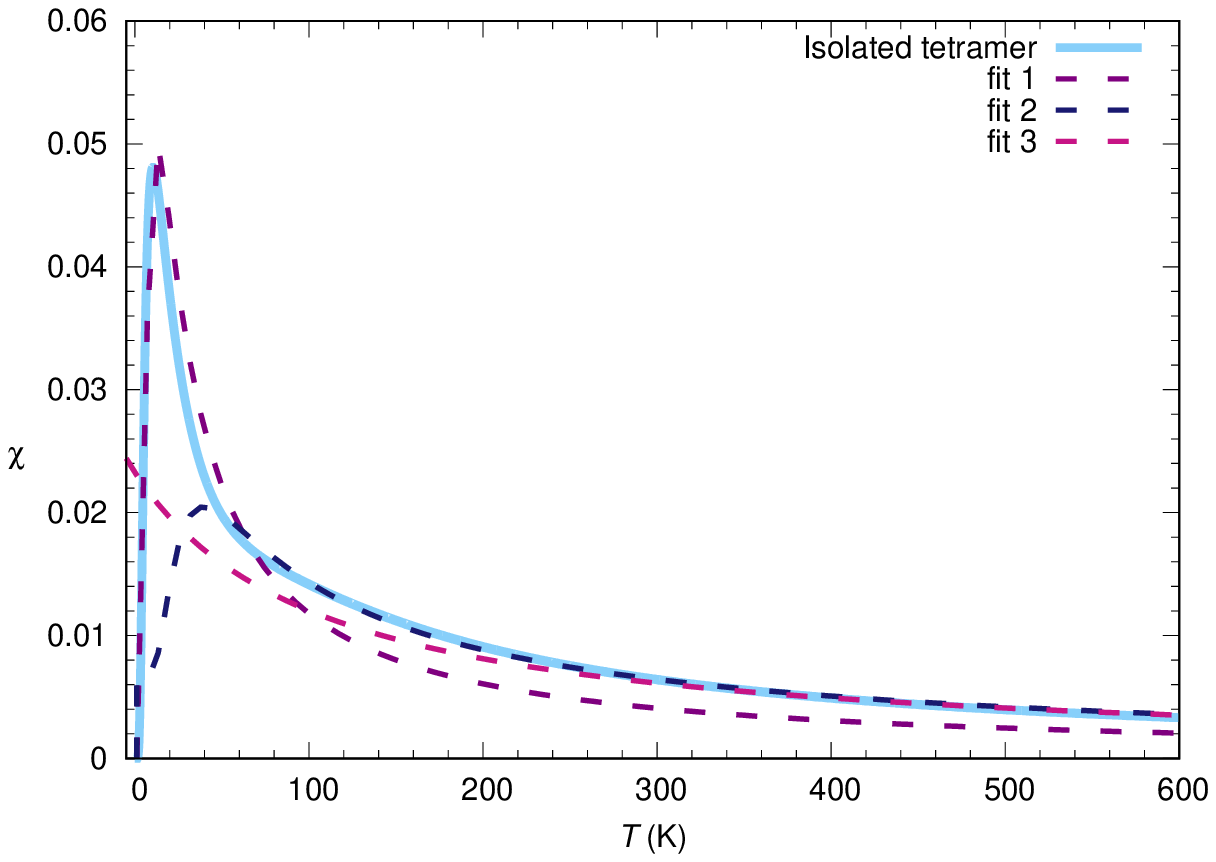}
		\caption{Magnetic susceptibility for a isolated tetramer. The dashed lines are best fits when number of spins are fixed in different regimes.}
	\end{figure}
	Susceptibility of a dimer spin system is give by (Ref. 42 in main text), 
	\begin{equation}
	\chi_{D}(T) = \frac{N_Dg^2\mu_B^2}{k_BT} \frac{e^{-J/K_BT}}{1+3e^{-J/k_BT}}.
	\end{equation}
	Susceptibility for antiferromagnetically or ferromagnetically correlated spins is given by the standard Curie-Weiss formula: 
	\begin{equation}
	\chi_{CW}(T) = \frac{N_{CW}g^2\mu_B^2S(S+1)}{3k_B}\frac{1}{T-T_{CW}}.
	\end{equation}
	In the above, $N_D$ is the number of spins forming singlets and $N_{CW}$ is the number of spins, $g$ is Lande $g$ factor , $k_B$ is Boltzmann constant, $\mu_B$ is Bohr magneton and $J$ is singlet-triplet energy gap.\\
	
	Susceptibility of an isolated tetramer is fitted using a combination of above two susceptibilities in different temperature regimes (see Table 1). Fitting parameters being $a_i=\frac{N_{D}g^2\mu_B^2}{k_B}$,  $b_i= \frac{J}{k_B}$, $c_i = \frac{N_{CW}g^2\mu_B^2S(S+1)}{3k_B}$ and $d_i = T_{CW}$. 
	For temperature region 1 $(0 K < T < 40 K)$, where the spins are expected to form a singlet. Fitting parameters reveal  $N_D \sim 1.61$ and $\frac{J}{k_B} \sim 16.79$. It is interesting to note that even though the system contains 4 spins the fit suggests the presence of a single dimer. This is because edge spins are strongly ferromagnetically coupled to the central spins that form a singlet, and  hence the tetramer effectively behaves like a singlet. Energy gap $J/k_B$ found from the fit also matches very well with first energy gap obtained from exact diagonalization. Fig. S4 (fit 1) highlights that the nature of system qualitatively remains same even if we set $N_D=2$ for temperature range $0 K < T < 40 K$. \\
	A combination of dimer and Curie-Weiss susceptibility was used in region 2 $(40 K < T < 300 K)$, with an expectation that the ferromagnetic coupling reduces, leaving a pure $\mathbf{S_2-S_3}$ singlet and two free spins. Number of free spins and spins involved in a dimer obtained from the fit confirms this picture.
	In Fig. S4, we illustrate that the fit is also reasonably good if we use $N_D = 2$ and $N_{CW} = 2$.
	For the higher temperature region $300 K < T < 600 K$ fit to Curie -Weiss susceptibility affirms the presence of free spins. Note the number of spins don't perfectly match due to the presence of finite but small coupling between all the spins, description in terms of these regions is only a simplified picture. Once again, if we use the simplified picture that all 4 spins in a tetramer contribute to Curie-Weiss behavior, the quality of the fit does not detriate much. Therefore, although in the main text we discuss an accurate fitting of the susceptibility data to a mixed dimer and Curie-Weiss behavior and identify three distinct regimes in temperature, here we show that even fixing $N_D$ and $N_{CW}$ to the naively expected values leads to good fits.
	
	\begin{table}
		\begin{center}
			\begin{tabular}{ | m{5em} | m{13em} | m{6em} | m{7em} | } 
				\hline
				& Fit Formula & \makecell{Fitting\\ parameters} & Inferences \\ 
				\hline
				\makecell{fit 1 \\ T=[0:40]} &  $ \chi(T) = \frac{a_1}{T} \frac{e^{-b_1/T}}{1+3e^{-b_1/T}}$ & \makecell{$a_1 = 4.02$ \\ $b_1 = 16.79$}  & \makecell{$N_D =1.61$\\ $J/k_B = 16.79$} \\ 
				\hline
				\makecell{fit 2 \\ T=[40:300]}&$\chi(T)=\frac{a_2}{T} \frac{e^{-b_2/T}}{1+3e^{-b_2/T}}+\frac{c_2}{T-d_2}$  & \makecell{$a_2 = 5.24 $ \\	$b_2 = 188.60$\\ $c_2 = 0.86$\\ $d_2 = 0.32$} & \makecell{$N_D = 2.1$\\ $J/k_B= 188.60$ \\ $N_{CW}= 1.38$ \\ $T_{CW}= 0.32$} \\ 
				\hline
				\makecell{fit 3 \\ T=[300:600]}  & $ \chi(T) = \frac{c_3}{T-d_3}$  & \makecell{$c_3 =2.06$\\ $d_3 = -21.19$} & \makecell{$N_{CW}= 3.31$\\ $T_{CW}= -21.19$} \\
				\hline
				
			\end{tabular}
			\caption{Fitting parameters of susceptibility for isolated tetramer. }
		\end{center}
	\end{table}
	
	\section*{Details of calculations for $N_c = 16$}
	
	The size of the Hilbert space for cluster Hamiltonian of a 16 spin cluster (four tetramers) is $2^{N_c} \sim 65000$. A brute force diagonalization of such large matrix multiple times to reach self consistency requires enormous computational time. Moreover, it turns out that if we allow for an unrestricted self-consistency approach wherein the mean-field vectors can point in any direction then the cluster Hamiltonian lacks many of the symmetries that are present in the full interacting Hamiltonian. For example, the bulk spins are not equivalent to edge spins and hence the translation symmetry is lost. The mean fields are not restricted to point along $z$ axis, leading to coupling terms of the form $S_1^+B_{1}^- + S_1^- B_{1}^+$ where $B^+_1$ and $B^-_1$ contain $x$ and $y$ components of mean fields acting on spin ${\bf S}_1$. The presence of these terms spoil conservation of total $S^z$. Therefore, the conservation of $z$-component of total spin does not hold in the general case.
	For $h_z=0$, to apply conservation of total spin ($S_{tot} ^z$) we restrict mean fields to be in $z$ direction. We divide the matrix into block diagonals with $S_{tot}^z= 0,1,2...8$ sectors. To further speed up calculation we only compute lowest 500 eigenstates in every sector. In order to justify this cut-off we show the comparison between the full eigen-spectrum and the truncated low-energy spectrum. The low energy eigen-spectrum is unaffected by the truncation (see Figure \ref{fig4}).
	
	\begin{figure}[ht]
		\centering
		\includegraphics[scale=0.7]{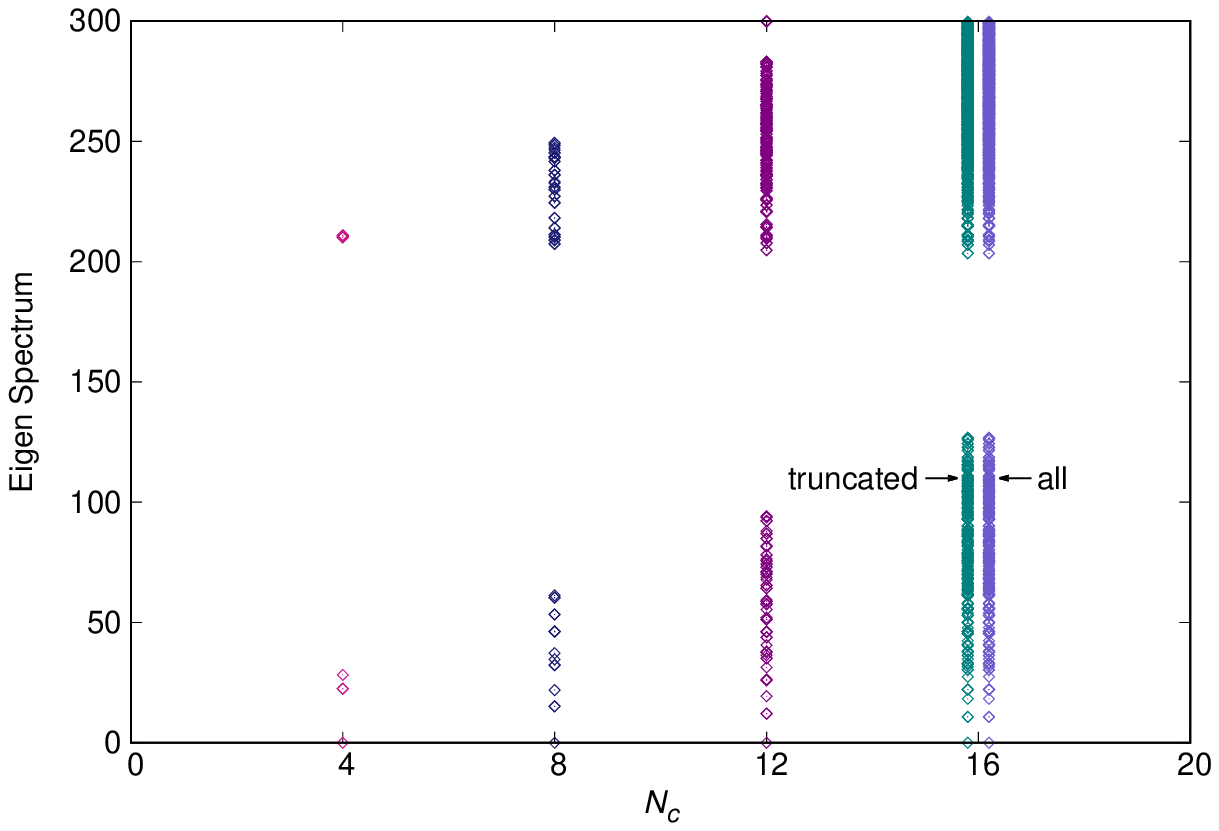}
		\caption{Eigen-spectrum for different values of $N_c$ shown as a tower of energies measured w.r.t. the ground state energy in each case. For $N_c = 16$ we show the comparison of the full spectrum with that of the low-energy spectrum consisting of 4500 states.}
		\label{fig4}
	\end{figure}

\end{document}